\documentclass[aps,pra,amsmath,amssymb,preprint]{revtex4-1}

\usepackage{dcolumn} 
\usepackage{bm}
\usepackage{graphicx}
\usepackage{longtable}
\usepackage{footnote}
\usepackage{float}

\usepackage{color}

\draft 

\begin{document}

\newcommand{\bra}[1]{\langle #1|}
\newcommand{\ket}[1]{|#1\rangle}
\newcommand{\braket}[2]{\langle #1|#2\rangle}
\newcommand{\p}{^\prime}
\newcommand{\pp}{^{\prime\prime}}

\title{Enhanced sensitivity to a possible variation of the \\proton-to-electron mass ratio in ammonia}

\author{A. Owens$^{1,2}$}\email{alec.owens.13@ucl.ac.uk}
\author{S. N. Yurchenko$^1$}
\author{W. Thiel$^2$}
\author{V. \v{S}pirko$^{3,4}$}
\affiliation{$^1$Department of Physics and Astronomy, University College London, Gower Street, WC1E 6BT London, United Kingdom}
\affiliation{$^2$Max-Planck-Institut f\"{u}r Kohlenforschung, Kaiser-Wilhelm-Platz 1, 45470 M\"{u}lheim an der Ruhr, Germany}
\affiliation{$^3$Academy of Sciences of the Czech Republic, Institute of Organic Chemistry and Biochemistry, Flemingovo n\'am.~2, 166 10 Prague 6, Czech Republic}
\affiliation{$^4$Department of Chemical Physics and Optics, Faculty of Mathematics and Physics, Charles University in Prague, Ke Karlovu 3, CZ-12116 Prague 2, Czech Republic}

\date{\today}

\begin{abstract}
Numerous accidental near-degeneracies exist between the $2\nu_2$ and $\nu_4$ rotation-vibration energy levels of ammonia. Transitions between these two states possess significantly enhanced sensitivity to a possible variation of the proton-to-electron mass ratio $\mu$. Using a robust variational approach to determine the mass sensitivity of the energy levels along with accurate experimental values for the energies, sensitivity coefficients have been calculated for over 350 microwave, submillimetre and far infrared transitions up to $J\!=\!15$ for $^{14}$NH$_3$. The sensitivities are the largest found in ammonia to date. One particular transition, although extremely weak, has a sensitivity of $T=-16,738$ and illustrates the huge enhancement that can occur between close-lying energy levels. More promising however are a set of previously measured transitions with $T=-32$ to $28$. Given the astrophysical importance of ammonia, the sensitivities presented here confirm that $^{14}$NH$_3$ can be used exclusively to constrain a spatial or temporal variation of $\mu$. Thus, certain systematic errors which affect the ammonia method can be eliminated. For all transitions analyzed we provide frequency data and Einstein A coefficients to guide future laboratory and astronomical observations.
\end{abstract}


\maketitle 

\section{Introduction}

  Molecules are an attractive testing ground for probing two particular dimensionless fundamental constants. Electronic transitions are sensitive to the fine structure constant $\alpha$, whilst vibration, rotation and inversion transitions are sensitive to the proton-to-electron mass ratio $\mu=m_p/m_e$. If any variation did exist it would manifest as observable shifts in the transition frequencies of certain molecular species. Such shifts can be detected by high-precision laboratory experiments over short time scales (years), or from astronomical observation of spectral lines at high redshift. The idea that the fundamental constants of nature may be understood within the framework of a deeper cosmological theory dates back to Dirac~\citep{Dirac:1937}. As of yet there is no theoretical justification for the values they assume, or even if they have always had the same values that we measure today. 
  
  Research in the field has become more active after claims of a temporal variation in the fine structure constant where observations of atomic absorption spectra of distant quasars suggested that $\alpha$ was smaller in the past~\citep{Webb:2001}. A few years later, measurements of H$_2$ spectra indicated that the proton-to-electron mass ratio was larger by $0.002\%$ up to twelve billion years ago~\citep{Reinhold:2006}. Numerous studies have followed and these have all produced null results (see Ref.~\citep{Ubachs:2016} for a detailed review). Any cosmological variation in the fundamental constants would require new physics beyond the Standard Model and as such, results are received with caution and must be confirmed, or refuted, with independent studies on different atomic and molecular absorbers.
  
  Ammonia has a large number of rotation-vibration transitions which are particularly sensitive to the proton-to-electron mass ratio~\citep{Jansen:2014,Spirko:2014,Owens:2015}. The sensitivity coefficient,
\begin{equation}
T_{u,l}=\frac{\mu}{E_u-E_l}\left(\frac{{\rm d}E_u}{{\rm d}\mu}-\frac{{\rm d}E_l}{{\rm d}\mu}\right)
\label{eq:T}
\end{equation}
where $E_u$ and $E_l$ is the energy of the upper and lower state, respectively, quantifies the effect that a possible variation of $\mu$ would have for a given transition. It is related to the frequency shift of the probed transition through the expression,
\begin{equation}
\frac{\Delta\nu}{\nu_0}=T_{u,l} \frac{\Delta\mu}{\mu_0}
\label{eq:shift}
\end{equation}
where $\Delta\nu=\nu_{\mathrm{obs}}-\nu_0$ is the change in the frequency and $\Delta\mu=\mu_{\mathrm{obs}}-\mu_0$ is the change in $\mu$, both with respect to their present day values $\nu_0$ and $\mu_0$. 

  The ammonia method~\citep{Flambaum:2007} (adapted from \citet{Veldhoven:2004}) compares inversion transitions in the vibrational ground state of $^{14}$NH$_3$ (henceforth referred to as NH$_3$) with rotational lines from other molecular species. By employing this approach several constraints on a temporal variation of $\mu$ have been established from measurements of the object B0218$+$357 at redshift $z\sim 0.685$~\citep{Flambaum:2007,Murphy:2008,Kanekar:2011}, and from the lensing galaxy PKS1830$-$211 at $z\sim 0.886$~\citep{Henkel:2009}. However, the reliance on other reference molecules, particularly those which are non-nitrogen bearing, is a major source of systematic error (see Refs.~\citep{Murphy:2008,Henkel:2009,Kanekar:2011} for a detailed discussion). 
  
  Methanol is now preferred because not only is it astronomically abundant, but it can be used exclusively to place limits on a drifting $\mu$~\citep{Jansen:2011,Levshakov:2011,Bagdonaite:2013a,Bagdonaite:2013b,Thompson:2013,Kanekar:2015}, circumventing the errors which arise from comparing different molecular species. The most robust constraint to date measured CH$_3$OH absorption lines in the system PKS1830$-$211~\citep{Kanekar:2015}. The three observed transitions possessed sensitivities ranging from $T=-1.0$ to $-7.4$ and were computed using an effective Hamiltonian model~\citep{Jansen:2011,Jansen:2011a}, by far the most common approach used for calculating sensitivity coefficients. 
  
  Despite being superseded by methanol, a comprehensive study of $^{14}$NH$_3$, $^{15}$NH$_3$, $^{14}$ND$_3$ and $^{15}$ND$_3$~\citep{Owens:2015} offered perspectives for the development of the ammonia method. Inversion transitions in the $\nu_4$ vibrational state had sensitivities from $T=-4.27$ to $4.67$, whilst the $^{14}$NH$_3$ astronomically observed $2_1^+\leftarrow 1_1^-$ and $0_0^-\leftarrow 1_0^+$ transitions in the $\nu_2$ state~\citep{Mauersberger:1988,Schilke:1992} possessed values of $T=17.24$ and $T=-6.59$, respectively. Here states are labelled as $J_{K}^{\pm}$ where $J$ is the rotational quantum number, $K$ is the projection onto the molecule-fixed $z$ axis, and $\pm$ denotes the parity of the state. Because of the abundance of NH$_3$ throughout the Universe and the ease with which its spectrum can be observed, identifying more transitions with large sensitivities in the microwave, submillimetre or far infrared regions could lead to a much tighter constraint on $\mu$. 

\begin{figure}
\includegraphics{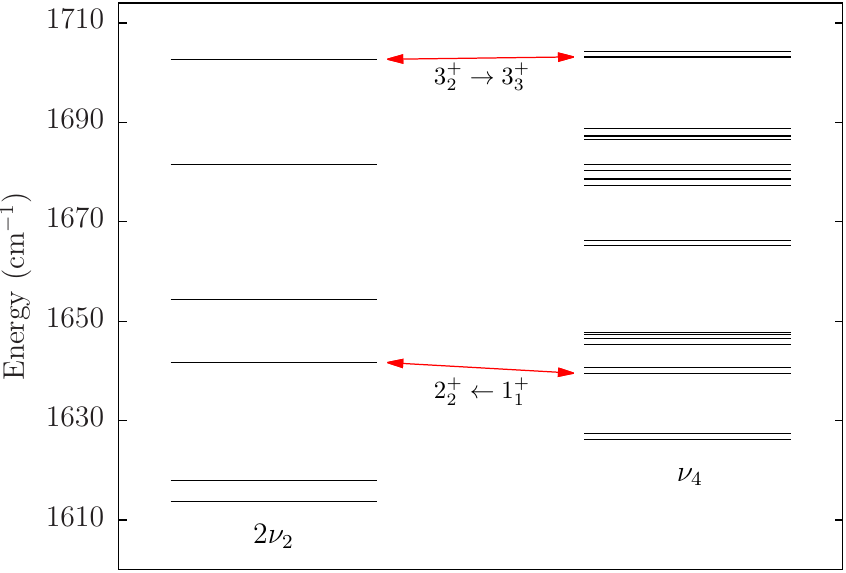}
\caption{\label{fig:degen_2v2_v4}Accidental near-degeneracies between the $2\nu_2$ and $\nu_4$ rotation-vibration energy levels of ammonia. Energy levels are labelled as $J_{K}^{\pm}$. For illustrative purposes only part of the rovibrational manifold is shown.}
\end{figure}
  
  A recent analysis of 56 sources of high-resolution $^{14}$NH$_3$ spectra utilizing the MARVEL procedure determined 4961 rovibrational energy levels of experimental quality, all labelled using a consistent set of quantum numbers~\citep{MARVEL:2015}. This has allowed us to investigate new transitions of NH$_3$ and accurately calculate their sensitivity to a possible variation of $\mu$. As shown in Fig.~\ref{fig:degen_2v2_v4}, numerous accidental near-degeneracies occur between the the $2\nu_2$ and $\nu_4$ rovibrational energy levels of ammonia. The strong Coriolis interaction between these two states~\citep{Urban:1980} can give rise to highly anomalous sensitivities. Furthermore, a large number of transitions between these levels have been measured experimentally~\citep{82SaHaAm.NH3,92SaEnHi.NH3}.
  
\section{Variational Approach}  

  To compute sensitivity coefficients the dependence on $\mu$ of the energy levels is required, i.e. the derivatives in Eq.~\eqref{eq:T}. Under the assumption that all baryonic matter may be treated equally~\citep{Dent:2007}, $\mu$ is assumed to be proportional to the molecular mass and it is sufficient to simply compute the mass dependence of the desired energy levels. The variational approach we employ here is identical to our previous study of ammonia and we refer the reader to Ref.~\citep{Owens:2015} (and references therein) for a detailed description. In short, a series of calculations are performed employing a scaled value for the mass of NH$_3$, from which numerical values of the derivatives ${\rm d}E/{\rm d}\mu$ are obtained by finite differences. After matching the derivatives with the experimentally determined energy levels from the MARVEL analysis, sensitivities are calculated through Eq.~\eqref{eq:T}. We also compute Einstein A coefficients to determine which transitions could realistically be detected. All calculations were carried out with the nuclear motion program TROVE~\citep{TROVE:2007}. Note that sensitivities have been computed for H$_3$O$^+$ and D$_3$O$^+$~\citep{15OwYuPo.H3Op} using exactly the same approach.

\section{Results and Discussion}  

\begin{figure}
\includegraphics{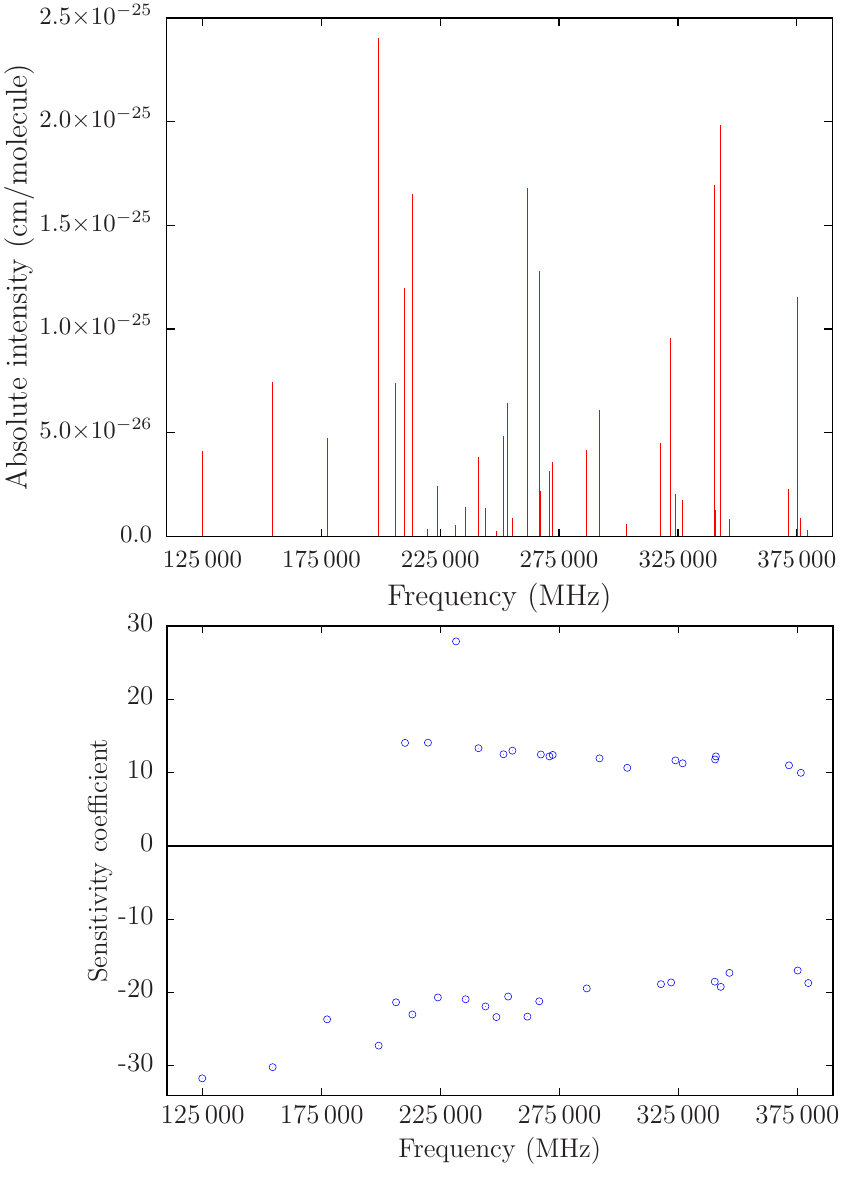}
\caption{\label{fig:observed_2v2_v4}Observed frequencies~\citep{82SaHaAm.NH3,92SaEnHi.NH3} and simulated intensities at temperature $T=296\,$K (top panel) with the corresponding sensitivities (bottom panel) for transitions between the $2\nu_2$ and $\nu_4$ vibrational states of NH$_3$.}
\end{figure}

  In Fig.~\ref{fig:observed_2v2_v4} we have simulated the intensities at room temperature for 38 previously observed transitions from Ref.~\citep{82SaHaAm.NH3,92SaEnHi.NH3} and plotted their corresponding sensitivity coefficients. The largest difference in sensitivity is $\Delta T=59.6$, which is over nine times more sensitive than the $\Delta T$ of the methanol lines used to establish the most robust constraint to date~\citep{Kanekar:2015}, and over seventeen times larger than the $\Delta T$ of the transitions utilized in the ammonia method~\citep{Flambaum:2007}. As well as being consistently large, the mixture of positive and negative sensitivities is highly beneficial for detecting a change in $\mu$ as transitions are shifted in opposing directions. From Fig.~\ref{fig:observed_2v2_v4}, one could imagine scanning this frequency window at two separate instances in time to produce a displaced spectrum if any variation of $\mu$ had occurred. In addition to the frequencies of Ref.~\citep{82SaHaAm.NH3,92SaEnHi.NH3}, there are 153 transitions with similar Einstein A coefficients and sensitivities from $T=-32.40$ to $17.27$ in the frequency range $100$ to $900\,$GHz. We provide comprehensive tables of all investigated transitions as supplementary material.
  
  The accuracy of the calculated sensitivity coefficients depends on the MARVEL energy levels and the computed TROVE numerical derivatives. The MARVEL analysis offers a rigorous evaluation of high-resolution NH$_3$ spectra. The $2\nu_2$ energy levels have an average error of $0.0027\,$cm$^{-1}$ for the 251 levels up to $J=15$, whilst a similar uncertainty of $0.0026\,$cm$^{-1}$ is given for the 495 $\nu_4$ energies. As such, the error on the predicted sensitivities is significantly reduced by replacing the computed TROVE energy levels with the corresponding MARVEL values in Eq.~\eqref{eq:T}. This is not to say that the TROVE frequencies are unreliable. As part of the MARVEL procedure the derived experimental energy levels are checked against theoretical predictions using the same potential energy surface (PES) and computational setup~\citep{Yurchenko:2011} as utilized for the present study. This PES is based on extensive high-level \textit{ab initio} calculations~\citep{05YuZhLi.NH3} and has subsequently been refined to experimental data up to $J\leq 8$~\citep{11YuBaTe.method}. If we calculate sensitivity coefficients for the 38 transitions shown in Fig.~\ref{fig:observed_2v2_v4} without replacing the energies, the TROVE sensitivities differ on average by $2.5\%$ to the MARVEL substituted sensitivities, the largest difference being $5.5\%$. Likewise for the additional 153 transitions with similar Einstein A coefficients, the TROVE sensitivities deviate on average by $2.1\%$. Such small differences reflect the quality of the underlying PES.
  
  However, the variational approach cannot account exactly for all near-degeneracies in the $2\nu_2$ and $\nu_4$ rovibrational manifold. A striking example of this is for the extremely weak $5_3^+(\nu_4)\leftarrow 5_2^+(2\nu_2)$ transition. A computed frequency of $\nu_{\mathrm{calc}}=3540.5\,$MHz has a sensitivity of $T_{\mathrm{calc}}=-1843.25$, already the largest known sensitivity coefficient for ammonia. Replacing with MARVEL energy levels gives $\nu_{\mathrm{exp}}=389.9\,$MHz and $T_{\mathrm{exp}}=-16,737.52$. The dramatic increase in magnitude occurs because of the inverse dependence on transition frequency (see Eq.~\eqref{eq:T}) and illustrates the huge enhancement that can happen between close-lying energy levels. Given the difference in predicted sensitivities one could question whether the computed numerical derivatives are still reliable. The change in frequency is just over $3000\,$MHz ($\approx0.1\,$cm$^{-1}$) so one would expect that they are reasonable. The difficulty is that quantifying the uncertainty of the numerical derivatives is not as straightforward because there are no analogous highly accurate experimental quantities.
  
  To investigate the error of the computed derivatives, new sensitivity coefficients were calculated using a purely \textit{ab initio} PES~\citep{05YuZhLi.NH3}. One can hope to establish a relationship between the difference in $\nu=E_u-E_l$, with the difference in the quantity ${\rm d}E_u/{\rm d}\mu-{\rm d}E_l/{\rm d}\mu$, by comparing values computed using this and the empirically refined PESs. Whilst no clear general correspondence between the uncertainty on these two quantities emerges, for near-coinciding energy levels separated by $1\,$cm$^{-1}$ or less, the percentage difference in ${\rm d}E_u/{\rm d}\mu-{\rm d}E_l/{\rm d}\mu$ is always smaller than the percentage difference in $\nu$. This ranges from 3-4 times smaller to several orders of magnitude smaller and suggests that for extremely close-lying energy levels, the underlying numerical derivatives are relatively stable. Thus, the huge amplification in sensitivity we predict is a result of replacing the theoretical frequencies with experimental values.
  
  For the transitions shown in Fig.~\ref{fig:observed_2v2_v4} and those with similar Einstein A coefficients, there is consistent agreement between the TROVE and MARVEL substituted sensitivities and errors in the computed derivatives will be negligible. When the two predictions differ significantly, which occurs for a number of weaker transitions with incredibly large sensitivities ranging from $T=-712.84$ to $509.21$ (see supplementary material), we are confident that the MARVEL substituted sensitivity coefficients are reliable. In all instances the residual between experiment and computed transition frequency never exceeds $1\,$cm$^{-1}$ (regarded as spectroscopic accuracy).
  
\section{Outlook}
    
  Finally, we briefly comment on possible experimental tests of our predictions. There are now novel techniques to produce ultracold polyatomic molecules~\citep{Zeppenfeld:2012}, which have rich spectra well suited for testing fundamental physics. Already experiments which decelerate, cool and trap ammonia molecules are being developed to probe a temporal variation of $\mu$~\citep{Veldhoven:2004,Sartakov:2008,Quintero:2013,Jansen:2013,Quintero:2014}. In Table~\ref{tab:weaker} we list several highly sensitive transitions, which despite being around two orders of magnitude weaker than the lowest intensity lines displayed in Fig.~\ref{fig:observed_2v2_v4}, could possibly be detected in such high-precision studies. 
  
\begin{table}
\tabcolsep=0.2cm
\caption{\label{tab:weaker}Highly sensitive weak transitions between the $2\nu_2$ and $\nu_4$ vibrational states of NH$_3$.}
\begin{center}
	\begin{tabular}{ccccc}
	\hline\hline
 	$\nu\p\leftarrow\nu\pp$ & $J_{K}^{\pm}\,\p\leftarrow J_{K}^{\pm}\,\pp$ & $\nu_{\mathrm{exp}}$/MHz & $A$/s$^{-1}$ & $T$ \\ 
	\hline
	$2\nu_2\leftarrow\nu_4$ & $2_{2}^{+}\leftarrow 1_{1}^{+}$ & 61\,712.7 & 1.042$\times 10^{-8}$ & 107.95 \\[-2mm]
	$\nu_4\leftarrow 2\nu_2$ & $7_{3}^{-}\leftarrow 7_{1}^{+}$ & 110\,957.2 & 9.461$\times 10^{-8}$ & -54.08 \\[-2mm]
	$\nu_4\leftarrow 2\nu_2$ & $10_{2}^{+}\leftarrow 10_{1}^{+}$ & 123\,427.8 & 4.745$\times 10^{-7}$ & -44.11 \\[-2mm]
	$\nu_4\leftarrow 2\nu_2$ & $6_{3}^{-}\leftarrow 6_{1}^{+}$ & 169\,341.3 & 2.539$\times 10^{-8}$ & -37.57 \\
    \hline\hline
    \end{tabular}
\end{center}
{\footnotesize All transitions are of symmetry $E\p\leftarrow E\pp$. Experimental frequencies have been obtained using energy levels from the MARVEL analysis~\citep{MARVEL:2015}}.
\end{table}
  
\begin{figure}
\includegraphics{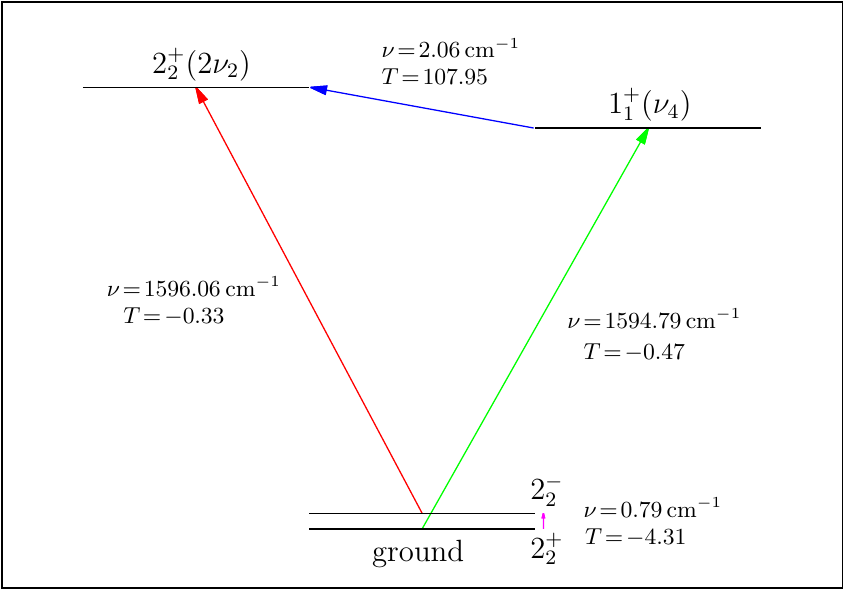}
\caption{\label{fig:CD}Use of combination differences involving infrared transitions from the ground vibrational state to the $2\nu_2$ and $\nu_4$ vibrational states of ammonia. Energy levels are labelled as $J_{K}^{\pm}$.}
\end{figure}

  If the transitions in Table~\ref{tab:weaker} are too weak to be detected directly, the use of combination differences involving infrared transitions from the ground vibrational state to the $2\nu_2$ and $\nu_4$ vibrational states should be considered. This technique would apply to any two levels provided transitions from a common ground state level can be identified, or for a situation such as that depicted in Fig.~\ref{fig:CD}. Infrared transitions to the respective levels of the $2_{2}^{+}(2\nu_2)\leftarrow 1_{1}^{+}(\nu_4)$ transition (sensitivity of $T=107.95$) have been measured experimentally~\citep{92SaEnHi.NH3}, whilst the corresponding ground state pure inversion frequency is well known~\citep{Lovas:2009}. Combination differences could be utilized to determine a possible shift in these energy levels provided the sensitivities of the three involved transitions are also known. The large number of potential combination differences prohibits us from carrying out a rigorous evaluation of all possible transitions. However, if particular combination differences could be readily measured in the future, it would be straightforward to compute the required sensitivity coefficients.
  
  Although laboratory experiments have greater control over systematic effects they provide only a local constraint on a drifting constant. It could be argued that even a null result would be limited to the age of the Solar System (around 4.6 billion years) and that a variation of $\mu$ could have occurred at earlier stages in the evolution of the Universe. More desirable are molecular systems which are astronomically relevant because observation at different redshifts presents the opportunity to look back to much earlier times in the Universe. Detection in a wide variety of cosmological settings also lends itself to searches for possible spatial variations of $\mu$, for which a number of studies using ammonia have been reported~\citep{Molaro:2009,Levshakov:2010b,Levshakov:2010a,Levshakov:2013}.
  
  The transitions presented in this study are perhaps more likely to be detected in terrestrial studies given that the rovibrational states involved lie above $1600\,$cm$^{-1}$. Astronomical detection is not impossible however. The energy levels of the $(J,K)=(18,18)$ inversion transition in the ground vibrational state of $^{14}$NH$_3$ reside at $2176.93\,$ and $2178.47\,$cm$^{-1}$, respectively, and this line was observed towards the galactic center star forming region Sgr B2~\citep{Wilson:2006}. A number of $2\nu_2\leftrightarrow\nu_4$ transitions which possess sizeable Einstein A coefficients and involve energy levels lower than the $(J,K)=(18,18)$ energies are listed in Table~\ref{tab:astro}. Such highly excited states could effectively be populated by exoergic chemical formation processes~\citep{Lis:2014}. It is hoped that future astronomical observations search for these particularly sensitive transitions.

\begin{table}
\tabcolsep=0.2cm
\caption{\label{tab:astro}Astronomically relevant transitions between the $2\nu_2$ and $\nu_4$ vibrational states of NH$_3$.}
\begin{center}
	\begin{tabular}{ccccc}
	\hline\hline
 	$\nu\p\leftarrow\nu\pp$ & $J_{K}^{\pm}\,\p\leftarrow J_{K}^{\pm}\,\pp$ & $\nu_{\mathrm{exp}}$/MHz & $A$/s$^{-1}$ & $T$ \\ 
	\hline
	$\nu_4\leftarrow 2\nu_2$ & $0_{0}^{+}\leftarrow 1_{1}^{+}$& 379\,596.5& 4.703$\times 10^{-6}$ & -18.70\\[-2mm]
	$\nu_4\leftarrow 2\nu_2$ & $1_{1}^{+}\leftarrow 1_{0}^{+}$& 824\,624.2& 6.427$\times 10^{-5}$& -9.13\\[-2mm]
	$2\nu_2\leftarrow \nu_4$ & $2_{1}^{+}\leftarrow 1_{0}^{+}$& 231\,528.2& 1.180$\times 10^{-6}$& 27.91\\[-2mm]
	$\nu_4\leftarrow 2\nu_2$ & $2_{2}^{+}\leftarrow 2_{1}^{+}$& 687\,852.5& 6.318$\times 10^{-5}$& -10.70\\[-2mm]
	$2\nu_2\leftarrow \nu_4$ & $3_{3}^{+}\leftarrow 2_{2}^{+}$& 489\,672.2& 4.360$\times 10^{-6}$& 12.72\\[-2mm]
	$\nu_4\leftarrow 2\nu_2$ & $3_{3}^{+}\leftarrow 3_{2}^{+}$& 557\,275.3& 7.623$\times 10^{-5}$& -12.87\\[-2mm]
	$2\nu_2\leftarrow \nu_4$ & $3_{2}^{+}\leftarrow 2_{1}^{+}$& 672\,644.4& 3.223$\times 10^{-5}$& 8.89\\[-2mm]
	$\nu_4\leftarrow 2\nu_2$ & $3_{2}^{+}\leftarrow 3_{1}^{+}$& 679\,163.4& 6.964$\times 10^{-5}$& -10.79\\[-2mm]
	$\nu_4\leftarrow 2\nu_2$ & $3_{1}^{+}\leftarrow 3_{0}^{+}$& 774\,889.5& 4.660$\times 10^{-5}$& -9.59\\[-2mm]
	$2\nu_2\leftarrow \nu_4$ & $3_{1}^{+}\leftarrow 2_{0}^{+}$& 842\,667.6& 1.210$\times 10^{-4}$& 6.91\\[-2mm]
	$\nu_4\leftarrow 2\nu_2$ & $4_{4}^{+}\leftarrow 4_{3}^{+}$& 441\,874.1& 7.796$\times 10^{-5}$& -15.68\\[-2mm]
	$\nu_4\leftarrow 2\nu_2$ & $4_{3}^{+}\leftarrow 4_{2}^{+}$& 548\,781.8& 9.102$\times 10^{-5}$& -12.94\\[-2mm]
	$\nu_4\leftarrow 2\nu_2$ & $4_{2}^{+}\leftarrow 4_{1}^{+}$& 657\,787.0& 6.078$\times 10^{-5}$& -11.07\\[-2mm]
	$\nu_4\leftarrow 2\nu_2$ & $5_{5}^{+}\leftarrow 5_{4}^{+}$& 342\,797.1& 7.054$\times 10^{-5}$& -19.22\\[-2mm]
	$\nu_4\leftarrow 2\nu_2$ & $5_{4}^{+}\leftarrow 5_{3}^{+}$& 434\,941.1& 9.782$\times 10^{-5}$& -15.59\\[-2mm]
	$\nu_4\leftarrow 2\nu_2$ & $5_{3}^{+}\leftarrow 5_{2}^{+}$ & 527\,333.3 & 8.219$\times 10^{-5}$ & -13.31\\[-2mm]
	$\nu_4\leftarrow 2\nu_2$ & $5_{2}^{+}\leftarrow 5_{1}^{+}$ & 618\,776.8 & 4.583$\times 10^{-5}$ & -11.66\\[-2mm]
	$\nu_4\leftarrow 2\nu_2$ & $5_{1}^{+}\leftarrow 5_{0}^{+}$ & 672\,376.5 & 2.542$\times 10^{-5}$ & -10.86\\[-2mm]
	$\nu_4\leftarrow 2\nu_2$ & $6_{6}^{+}\leftarrow 6_{5}^{+}$ & 261\,535.4 & 5.745$\times 10^{-5}$ & -23.29\\[-2mm]
	$\nu_4\leftarrow 2\nu_2$ & $6_{5}^{+}\leftarrow 6_{4}^{+}$ & 340\,322.9 & 9.137$\times 10^{-5}$ & -18.52\\[-2mm]
	$\nu_4\leftarrow 2\nu_2$ & $6_{4}^{+}\leftarrow 6_{3}^{+}$ & 413\,748.1 & 9.006$\times 10^{-5}$ & -15.98\\[-2mm]
	$\nu_4\leftarrow 2\nu_2$ & $6_{3}^{+}\leftarrow 6_{2}^{+}$ & 488\,661.3 & 6.308$\times 10^{-5}$ & -14.12\\[-2mm]
	$\nu_4\leftarrow 2\nu_2$ & $6_{2}^{+}\leftarrow 6_{1}^{+}$ & 559\,214.0 & 3.027$\times 10^{-5}$ & -12.73\\[-2mm]
	$\nu_4\leftarrow 2\nu_2$ & $7_{7}^{+}\leftarrow 7_{6}^{+}$ & 198\,997.4 & 4.284$\times 10^{-5}$ & -27.24\\[-2mm]
	$\nu_4\leftarrow 2\nu_2$ & $7_{6}^{+}\leftarrow 7_{5}^{+}$ & 266\,541.0 & 7.700$\times 10^{-5}$ & -21.20\\[-2mm]
	$\nu_4\leftarrow 2\nu_2$ & $7_{5}^{+}\leftarrow 7_{4}^{+}$ & 321\,935.0 & 8.437$\times 10^{-5}$ & -18.62\\[-2mm]
	$\nu_4\leftarrow 2\nu_2$ & $7_{4}^{+}\leftarrow 7_{3}^{+}$ & 375\,174.5 & 6.887$\times 10^{-5}$ & -16.99\\[-2mm]
	$\nu_4\leftarrow 2\nu_2$ & $7_{3}^{+}\leftarrow 7_{2}^{+}$ & 430\,468.6 & 4.113$\times 10^{-5}$ & -15.29\\[-2mm]
	$\nu_4\leftarrow 2\nu_2$ & $8_{8}^{+}\leftarrow 8_{7}^{+}$ & 154\,415.5 & 3.036$\times 10^{-5}$ & -30.19\\
    \hline\hline
    \end{tabular}
\end{center}
{\footnotesize For symmetry of transitions see supplementary material. Experimental frequencies from Ref.~\citep{82SaHaAm.NH3,92SaEnHi.NH3} or obtained using energy levels from the MARVEL analysis~\citep{MARVEL:2015}.}
\end{table}

\begin{acknowledgments}
S.Y. thanks ERC Advanced Investigator Project 267219. V.S. acknowledges research project RVO:61388963 (IOCB) and support from the Czech Science Foundation (grant P209/15-10267S).
\end{acknowledgments}

\clearpage
\newpage
\bibliographystyle{apsrev}

\end{document}